\documentclass[twocolumn,showpacs,floats,floatfix,superscriptaddress,aps,pra]{revtex4}
\usepackage{amsfonts}
\usepackage{amssymb}
\usepackage{amsmath}
\usepackage{graphicx}
\usepackage{bm}
\usepackage{color}
\usepackage{epsfig}
\usepackage{calc}
\usepackage{ifthen}

\begin{document}

\title{Complete achromatic optical switching between two waveguides with a
sign flip of the phase mismatch}
\date{\today }

\begin{abstract}
We present a two-waveguide coupler which, realizes complete achromatic
all-optical switching. The coupling of the waveguides has a
hyperbolic-secant shape while the phase mismatch has a sign flip at the
maximum of the coupling. We derive an analytic solution for the electric
field propagation using coupled mode theory and show that the light
switching is robust again small-to-moderate variations in the coupling and
phase mismatch. Thus, we realize an \emph{achromatic light switching}
between the two waveguides. We further consider the extended case of three
coupled waveguides in an array and pay special attention to the case of
equal achromatic light beam splitting.
\end{abstract}

\author{Wei Huang}
\affiliation{Engineering Product Development, Singapore University of Technology and
Design, 20 Dover Drive, 138682 Singapore}
\author{Andon A. Rangelov}
\affiliation{Engineering Product Development, Singapore University of Technology and
Design, 20 Dover Drive, 138682 Singapore}
\affiliation{Department of Physics, Sofia University, James Bourchier 5 blvd., 1164
Sofia, Bulgaria}
\author{Elica Kyoseva}
\affiliation{Engineering Product Development, Singapore University of Technology and
Design, 20 Dover Drive, 138682 Singapore}
\maketitle



\section{Introduction}

\label{Introduction}

The spatial light propagation in engineered coupled waveguide arrays is of
fundamental importance to wave optics \cite{Yariv73,Haus,Yariv90}. The
electric field propagation in waveguide arrays can be accurately described
within the coupled mode theory \cite{Yariv73,Haus,Yariv90} and the resulting
optical wave equation describing the spatial propagation of monochromatic
light in dielectric structures is remarkably similar to the temporal Schr%
\"{o}dinger equation describing a quantum-optical system driven by an
external electromagnetic field \cite{Longhi09}. The simplest realization of
a waveguide array consists of two identical evanescently-coupled parallel
waveguides. In this case, light is periodically switched between the
waveguides throughout the evolution \cite{Yariv90} in analogy to the
quantum-optical Rabi oscillations \cite{LonghiPRA05}. Consequently, more
complex waveguide configurations were designed to realize rich physical
phenomena and for several of them analytical solutions for the light
propagation have been described in the literature \cite%
{Longhi05,Ornigotti,DreisowPRA}.

In this work, we study the optical switching between two evanescently
coupled planar waveguides whose coupling has a hyperbolic-secant shape and
the phase mismatch is constant with a sign flip at the coupling maximum. We
derive an analytic solution for complete light transfer (CLT) and we show
that CLT is robust against variations in the experimental parameters;
therefore, the technique is expected to find applications in \textit{%
achromatic light switching}. Furthermore, we extend the model to three
evanescently coupled waveguides in planar array and show that starting from
the middle waveguide light\ can be equally split between the outer ones. We
show that the light splitting is insensitive to fluctuations in the coupling
and phase mismatch of the waveguides. Hence, this set-up may find an
important technological application as an \textit{achromatic light
beam-splitter}. It is important to note, that the coupling model, which we
consider here bears a close connection with the phase jump models from
quantum optics \cite{Vitanov,Torosov}, where the phase jump is instead in
the coupling rather than the detuning. Such a model would also realize CLT
in the system of two coupled waveguides but engineering a sign flip in the
coupling would be a significant technological challenge.


\section{Model of two coupled waveguides}

\label{Model of two coupled waveguides}

\begin{figure}[tbh]
\centerline{\includegraphics[width=0.6\columnwidth]{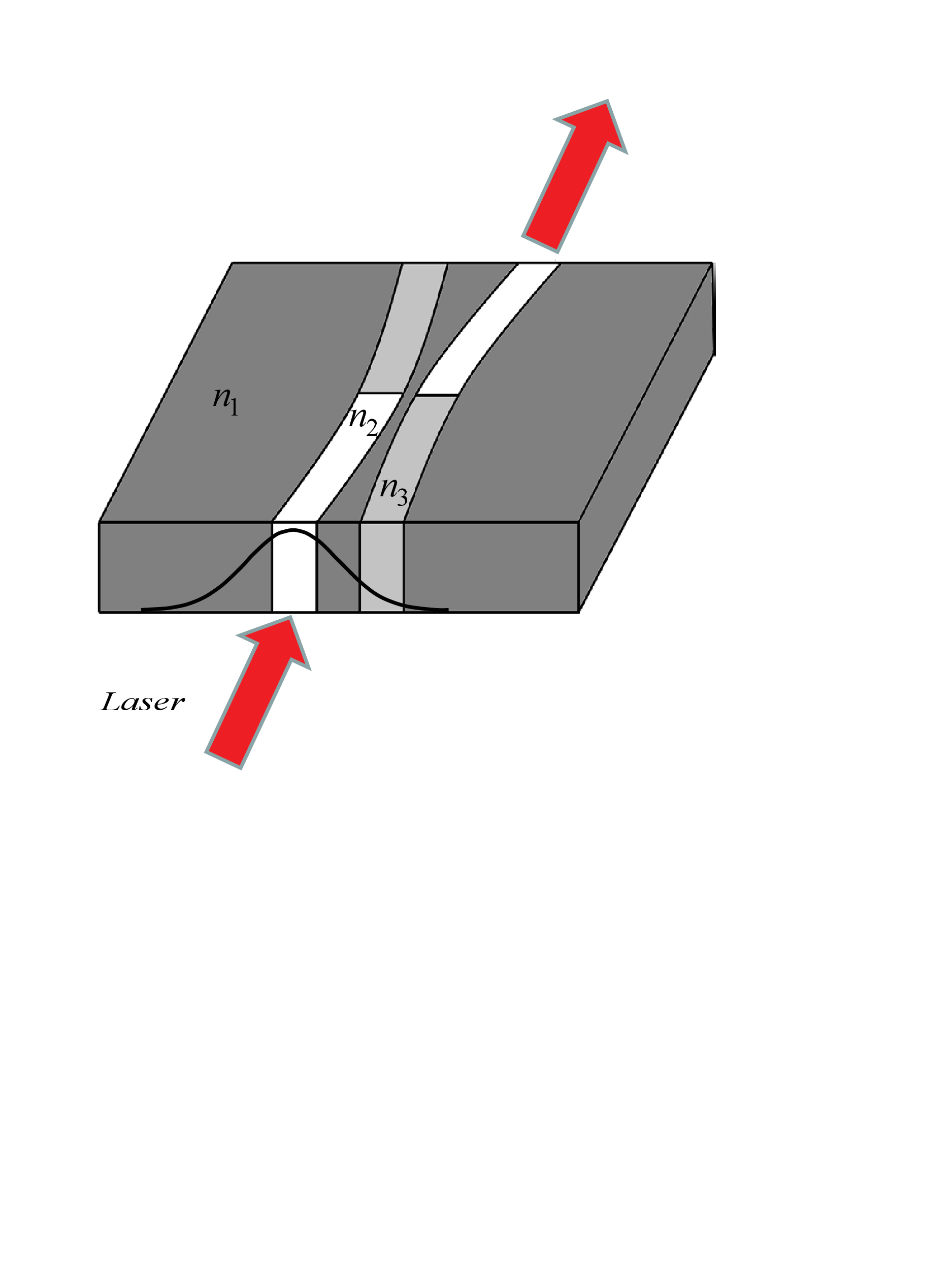}}
\caption{(Color online) Evanescently coupled two waveguides made of two
slabs with refractive indexes $n_{2}$ and $n_{3}$, embedded in a medium with
an index of refraction $n_{1}$. Gaussian shaped beam light is injected
initially in the left waveguide and at the end of the adiabatic evolution
the light is achromatically switched to the right waveguide.}
\label{Fig1}
\end{figure}

We consider two evanescently-coupled planar optical waveguides as shown in
Fig. \ref{Fig1}. In the paraxial approximation, the propagation of a
monochromatic light beam in the waveguides can be analyzed in the framework
of the coupled mode theory (CMT) \cite{Yariv73,Haus,Yariv90}. The
corresponding evolution of the wave amplitudes can be described by a set of
two coupled differential equations (in matrix form),
\begin{equation}
i\dfrac{d}{dz}C(z)=\mathbf{H}(z)C(z),  \label{Schrodinger equation}
\end{equation}%
which has the form of a Schr\"{o}dinger equation \cite{Longhi09} with $z$
being the longitudinal coordinate. The components of the vector $C(z)=\left[
c_{1}(z),c_{2}(z)\right] ^{\mathnormal{T}}$ are the amplitudes of the
fundamental modes in the two waveguides and $I_{1,2}=\left\vert
c_{1,2}(z)\right\vert ^{2}$ are the corresponding normalized light
intensities. The operator $\mathbf{H}(z)$ describes the interaction between
the waveguide modes and is explicitly given as,
\begin{equation}
\mathbf{H}(z)=%
\begin{bmatrix}
\beta _{1}(z) & \Omega (z) \\
\Omega (z) & \beta _{2}(z)%
\end{bmatrix}%
.  \label{Hamiltonian}
\end{equation}%
Here, $\beta _{k}(z)$ with $k=(1,2)$ is the constant propagation coefficient
of the $k$-th waveguide and $\Omega (z)$ is the variable coupling
coefficient between the waveguides. We note that only the difference between
the diagonal terms $\Delta (z)=\beta _{1}(z)-\beta _{2}(z)$ is important and
it is called phase mismatch. We remove $\beta _{1}(z)$ and $\beta _{2}(z)$
from Eq. (\ref{Hamiltonian}) by incorporating them as phases in the
amplitudes $c_{1}(z)$ and $c_{2}(z)$. Hence, Eq. (\ref{Schrodinger equation}%
) obtains the following form,
\begin{equation}
i\dfrac{d}{dz}\left[
\begin{array}{c}
c_{1}(z) \\
c_{2}(z)%
\end{array}%
\right] =%
\begin{bmatrix}
-\Delta (z) & \Omega (z) \\
\Omega (z) & \Delta (z)%
\end{bmatrix}%
\left[
\begin{array}{c}
c_{1}(z) \\
c_{2}(z)%
\end{array}%
\right] .  \label{two-state atom}
\end{equation}

In the next section we shall derive the solution to Eq. (\ref{two-state atom}%
) for the step-sech model for which the coupling $\Omega (z)$ and phase
mismatch $\Delta (z)$ are given by
\begin{subequations}
\label{step-sech model}
\begin{eqnarray}
\Omega (z) &=&\Omega _{0}\,\text{sech}\left( z/L\right) , \\
\Delta (z) &=&\left\{
\begin{array}{c}
\Delta _{0} \\
-\Delta _{0}%
\end{array}%
\right.
\begin{array}{c}
(z<0) \\
(z>0)%
\end{array}%
.
\end{eqnarray}
Here, $L$ is the full width at half maximum for the coupling $\Omega (z)$
and we have also chosen the point $z=0$ to be the middle of the waveguides.
Without loss of generality, the constants $\Omega _{0}$ and $\Delta _{0}$
are assumed positive.


\section{Exact analytical solution for the spatial light propagation}

\label{Sec3}


In order to derive the evolution of the wave amplitudes we rewrite Eq. (\ref%
{two-state atom}) into the interaction picture,
\end{subequations}
\begin{equation}
i\dfrac{d}{dz}\left[
\begin{array}{c}
c_{1}(z) \\
c_{2}(z)%
\end{array}%
\right] =%
\begin{bmatrix}
0 & \Omega (z)e^{-iD\left( z\right) } \\
\Omega (z)e^{iD\left( z\right) } & 0%
\end{bmatrix}%
\left[
\begin{array}{c}
c_{1}(z) \\
c_{2}(z)%
\end{array}%
\right] ,  \label{interaction picture}
\end{equation}%
with $D(z)=\int_{z_{i}}^{z_{f}}\Delta (z)dz$. Then, we can decouple $%
c_{1}(z) $ from $c_{2}(z)$ by taking a second derivative in $z$, which gives
\begin{equation}
\dfrac{d^{2}}{dz^{2}}c_{1}(z)-\left( \frac{\dot{\Omega}}{\Omega }-i\Delta
\right) \dfrac{d}{dz}c_{1}(z)+\Omega ^{2}c_{1}(z)=0.
\label{second order differential equation}
\end{equation}%
The next step is to change the independent variable from $z$ to $t(z)=\tfrac{%
1}{2}[1+$\textnormal{tanh}$(z/L)]$ noting that $t(-\infty )=0$, $t(0)=\tfrac{%
1}{2}$ and $t(+\infty )=1$. We thus rewrite Eq. (\ref{second order
differential equation}) for $c_{1}(t(z))$ using Eqs. (\ref{step-sech model})
as,
\begin{equation}
t(1-t)\dfrac{d^{2}c_{1}}{dt^{2}}+(\tfrac{1}{2}-t+\tfrac{1}{2}i\Delta _{0}L)%
\dfrac{dc_{1}}{dt}+\Omega _{0}^{2}L^{2}c_{1}=0.
\label{Gauss Hypergeometric equation}
\end{equation}%
This equation has the same form as the Gauss Hypergeometric equation \cite%
{Erdelyi,Abramowitz}:
\begin{equation}
t(1-t)\ddot{x}+[\gamma -(\alpha +\beta +1)t]\dot{x}-\alpha \beta x=0,
\label{Gauss}
\end{equation}%
where the overdot denotes a derivative in $t$. The solution to Eq. %
\eqref{Gauss} is given in terms of a linear combination of two Gauss
Hypergeometric functions, $F(\alpha ,\beta ,\gamma ;t)$ and $t^{1-\gamma
}F(\alpha +1-\gamma ,\beta +1-\gamma ,2-\gamma ;t)$ \cite%
{Erdelyi,Abramowitz,Vitanov,Torosov}. For the system considered here and
described by Eq. (\ref{Gauss Hypergeometric equation}) the corresponding
parameters are
\begin{equation}
\alpha =\Omega _{0}L,\quad \beta =-\alpha ,\quad \gamma =\tfrac{1}{2}+\tfrac{%
1}{2}i\Delta _{0}L.  \label{alpha}
\end{equation}

We then find that the solution to Eq. (\ref{Gauss Hypergeometric equation})
is given by,
\begin{eqnarray}
c_{1}(t) &=&AF(\alpha ,-\alpha ,\gamma ;t)+  \notag \\
&&Bt^{1-\gamma }F(\alpha +1-\gamma ,1-\alpha -\gamma ,2-\gamma ;t),
\end{eqnarray}%
where $A$ and $B$ are integration constants. Furthermore, using $%
c_{2}(t)=(ie^{i\Delta t}\dot{c}_{1}(t)\dfrac{dt}{dz})/\Omega (t)$ for the
second amplitude we obtain,
\begin{widetext}
\begin{equation}
c_{2}(t)=i(1-t)^{1-\gamma }[-A\dfrac{\alpha }{\gamma }t^{\gamma }F(1+\alpha,1-\alpha ,1+\gamma ,t)+B\dfrac{1-\gamma }{\alpha }F(1 + \alpha -\gamma, 1- \alpha - \gamma ,1 - \gamma; t)].
\end{equation}
\end{widetext}The integration constants $A$ and $B$ depend on the initial
conditions and are given by
\begin{eqnarray}
A &=&c_{1}(0), \\
B &=&-i\alpha \dfrac{c_{2}(0)}{(1-\gamma )}.
\end{eqnarray}%
Hence, for $0\leq t\leq \frac{1}{2}$ ($-\infty <z\leq 0$), the wave
amplitudes evolve according to $C(t)=\mathbf{U}(t,0)C(0)$ with
\begin{gather}
U_{11}(t,0)=U_{22}^{\star }(t,0)=F(\alpha ,-\alpha ,\gamma ;t), \\
U_{12}(t,0)=-U_{21}^{\star }(t,0)=  \notag \\
=\dfrac{-i\alpha }{1-\gamma }t^{1-\gamma }F(\alpha +1-\gamma ,1-\alpha
-\gamma ,2-\gamma ;t).
\end{gather}%
That is, the propagator from $z\rightarrow -\infty $ ($t=0$) to $z=0$ ($t=%
\frac{1}{2}$) can be simply expressed as,
\begin{equation}
\mathbf{U}(\tfrac{1}{2},0)=%
\begin{bmatrix}
a & -b^{\star } \\
b & a^{\star }%
\end{bmatrix}%
,
\end{equation}%
where
\begin{eqnarray}
a &=&F(\alpha ,-\alpha ,\gamma ;\frac{1}{2})=\frac{\pi ^{1/2}}{2^{\gamma }}%
\Gamma \left( \gamma \right) \left( \xi +\eta \right) , \\
b &=&-\dfrac{i\alpha }{2\gamma }F(1+\alpha ,1-\alpha ,1+\gamma ;\frac{1}{2})
\notag \\
&=&\frac{-i\pi ^{1/2}}{2^{\gamma }}\Gamma \left( \gamma \right) \left( \xi
-\eta \right) ,
\end{eqnarray}%
with the exact form of the parameters
\begin{eqnarray}
\xi &=&\frac{1}{\Gamma (\tfrac{1}{4}+\tfrac{1}{2}\alpha +\tfrac{1}{4}i\Delta
_{0}L)\Gamma (\tfrac{3}{4}-\tfrac{1}{2}\alpha +\tfrac{1}{4}i\Delta _{0}L)},
\\
\eta &=&\frac{1}{\Gamma (\tfrac{3}{4}+\tfrac{1}{2}\alpha +\tfrac{1}{4}%
i\Delta _{0}L)\Gamma (\tfrac{3}{4}-\tfrac{1}{2}\alpha +\tfrac{1}{4}i\Delta
_{0}L)}.
\end{eqnarray}%
Using the time-symmetry of equation (\ref{two-state atom}) and taking into
account that the only change for $z>0$ is the sign of $\Delta \left(
z\right) $, it is a simple matter to show that the propagator for $0\leq
z<\infty $ ($\tfrac{1}{2}\leq t\leq 1$) reads
\begin{equation}
\mathbf{U}(1,\tfrac{1}{2})=%
\begin{bmatrix}
a & -b \\
b^{\star } & a^{\star }%
\end{bmatrix}%
.
\end{equation}%
The total evolution propagator $\mathbf{U}(1,0)=\mathbf{U}(1,\tfrac{1}{2})%
\mathbf{U}(\tfrac{1}{2},0)$ is
\begin{equation}
\mathbf{U}(1,0)=%
\begin{bmatrix}
a^{2}-b^{2} & -2\,\Re (ab^{\star }) \\
2\,\Re (ab^{\star }) & (a^{2}-b^{2})^{\star }%
\end{bmatrix}%
.
\end{equation}

Then the normalized light intensity in the second waveguide is given by
\begin{equation}
I_{2}=|U_{12}(1,0)|^{2}=|2\,\Re (ab^{\star })|^{2},  \label{intensity}
\end{equation}

Finally, we obtain the analytical expression for the light transfer between
waveguides by substituting $a$, $b$, $\xi $ and $\eta $ in Eq. (\ref%
{intensity})
\begin{equation}
I_{2}=\left[ \text{sech}\left( \frac{\pi \Delta _{0}L}{2}\right) \Im \left(
e^{i\phi }\cos \left( \pi \alpha +\frac{i\pi \Delta _{0}L}{2}\right) \right) %
\right] ^{2},
\end{equation}%
where%
\begin{equation}
\phi =2\arg \left[ \Gamma (\tfrac{1}{4}-\tfrac{1}{2}\alpha -\tfrac{1}{4}%
i\Delta _{0}L)\Gamma (\tfrac{1}{4}+\tfrac{1}{2}\alpha +\tfrac{1}{4}i\Delta
_{0}L)\right] .
\end{equation}%
Recalling equation (\ref{alpha}) and using the asymptotic expansions of $\ln
\left( \Gamma \right) $ \cite{Erdelyi,Abramowitz} in the limit of large
coupling ($\Omega _{0}>>\Delta _{0}$) the light intensity in the second
waveguide turns to
\begin{equation}
I_{2}\approx \frac{\Omega _{0}^{2}}{\Omega _{0}^{2}+\Delta _{0}^{2}}\left[ 1-%
\frac{2\Delta _{0}e^{-\pi \Delta _{0}L/2}}{\Omega _{0}}\cos \left( \tfrac{1}{%
2}\pi \Omega _{0}L\right) +\mathcal{O}^{2}\right] ^{2},
\end{equation}%
hence \emph{complete light switching} between the two waveguides occurs.


\section{Adiabatic solution}

\label{Sec4}


We shall now derive the adiabatic solution for the general model where the
waveguides' coupling is a symmetric pulse-shaped smooth function and the
phase mismatch has a sign flip at the coupling maximum.

\begin{figure}[t]
\includegraphics[width=0.8\columnwidth]{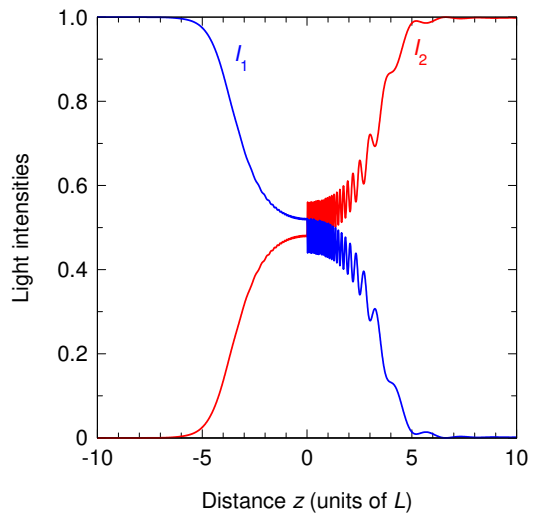}
\caption{(Color online) Demonstration of complete light switching between
two waveguides. We assume sech-shaped coupling $\Omega (z)=\Omega _{0}\,$sech%
$\left( z/L\right) $ and step-phase mismatch $\Delta (z)=\Delta _{0}\,\left(
1-2\protect\theta (z)\right) $, $\protect\theta (z)$ is being the Heaviside
step function. We select adiabatic parameters $\Omega _{0}=50/L$ and $\Delta
_{0}=2/L$. }
\label{Fig2}
\end{figure}

First we write Eq. (\ref{Schrodinger equation}) in the adiabatic basis \cite%
{Allen,Shore,Vitanov 2001}
\begin{equation}
i\dfrac{d}{dz}%
\begin{bmatrix}
a_{1}(z) \\
a_{2}(z)%
\end{bmatrix}%
=%
\begin{bmatrix}
-\varepsilon (z) & -i\dot{\theta}(z) \\
i\dot{\theta}(z) & \varepsilon (z)%
\end{bmatrix}%
\begin{bmatrix}
a_{1}(z) \\
a_{2}(z)%
\end{bmatrix}%
,  \label{adiabatic basis}
\end{equation}%
where the overdot denotes a derivative in the longitudinal coordinate $z$ and%
\begin{eqnarray}
\varepsilon (z) &=&\sqrt{\Omega ^{2}(z)+\Delta ^{2}(z)}, \\
\tan \left( 2\theta \right) &=&\frac{\Omega (z)}{\Delta (z)}.
\end{eqnarray}%
The amplitudes $a_{1}(z)$ and $a_{2}(z)$ in the adiabatic basis are
connected with the diabatic (original) ones, $c_{1}(z)$ and $c_{2}(z)$, via
the rotation matrix
\begin{equation}
\mathbf{R}\left( \theta \left( z\right) \right) =%
\begin{bmatrix}
\cos \theta & \sin \theta \\
-\sin \theta & \cos \theta%
\end{bmatrix}%
,
\end{equation}%
as $\left( c_{1}(z),c_{2}(z)\right) ^{T}=\mathbf{R}\left( \theta \left(
z\right) \right) \left( a_{1}(z),a_{2}(z)\right) ^{T}$. When the evolution
of the system is adiabatic, $\left\vert a_{1}(z)\right\vert $ and $%
\left\vert a_{2}(z)\right\vert $ remain constant \cite{Allen,Shore,Vitanov
2001}. Mathematically, adiabatic evolution means that the non-diagonal terms
in Eq. (\ref{adiabatic basis}) are small compared to the diagonal terms and
can be neglected. This restriction amounts to the following adiabatic
condition on the interaction parameters \cite{Allen,Shore,Vitanov 2001}:
\begin{equation}
\left\vert \dot{\Omega}\Delta -\dot{\Delta}\Omega \right\vert \ll \left\vert
\Omega ^{2}+\Delta ^{2}\right\vert ^{3/2}.
\end{equation}%
When the evolution is adiabatic the solution for the propagator in the
adiabatic basis from an initial coordinate $z_{i}$ to a final coordinate $%
z_{f}$ reads
\begin{equation}
\mathbf{U}^{\mathnormal{ad}}\left( z_{f},z_{i}\right) =%
\begin{bmatrix}
\exp \left( -iS\right) & 0 \\
0 & \exp \left( iS\right)%
\end{bmatrix}%
,
\end{equation}%
where $S=\int_{z_{i}}^{z_{f}}\sqrt{\Omega ^{2}(z)+\Delta ^{2}(z)}dz$. The
full propagator in the original basis for the model given in Eq. (\ref%
{step-sech model}) reads
\begin{widetext}
\begin{equation}
\mathbf{U} (z_{f},z_{i}) =\mathbf{R} ( \theta ( z_{f} )) \mathbf{U} ( z_{f},0) \mathbf{R}^{-1} (\theta ( z\rightarrow +0)) \mathbf{R} ( \theta (z \rightarrow -0 )) \mathbf{U} (0,z_{i}) \mathbf{R}^{-1} (\theta ( z_{i} )) .
\end{equation}
\end{widetext}

\begin{figure}[tbh]
\centerline{\includegraphics[width=0.90\columnwidth]{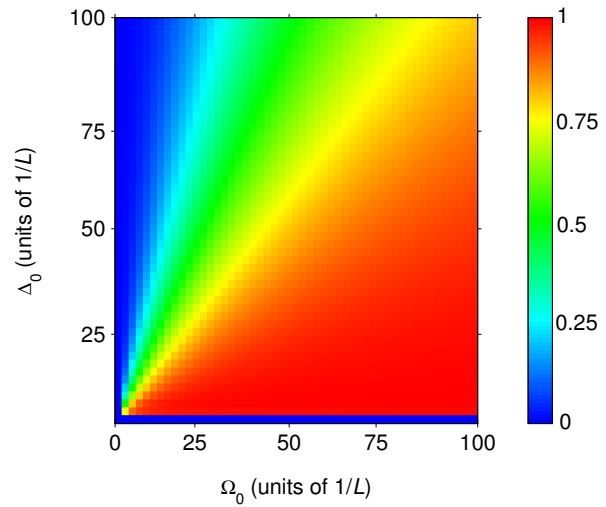}}
\caption{(Color online) Contour plot of the final intensity in the right
waveguide (waveguide two) from Fig. \protect\ref{Fig1}, if initially the
light is injected only into the left waveguide (waveguide one). Numerically
simulated contour plot is made for Eq. (\protect\ref{two-state atom}) using
the coupling $\Omega _{0}$ and the phase mismatch $\Delta _{0}$ from Eq. (%
\protect\ref{step-sech model}), with the longitudinal coordinate $z$ that
change from $-10L$ to $10L$.}
\label{Fig3}
\end{figure}

Therefore if we take into account that $\Omega (z_{i})=\Omega (z_{f})=0$ and
$\Delta (z\rightarrow -0)=-\Delta (z\rightarrow +0)=\Delta _{0}$ then the
light intensity transfer to the second waveguide is
\begin{equation}
I_{2}(z_{f})=\left\vert \mathbf{U}_{12}\left( z_{f},z_{i}\right) \right\vert
^{2}=\frac{\Omega _{0}^{2}}{\Omega _{0}^{2}+\Delta _{0}^{2}}.
\label{complete light transfer}
\end{equation}%
Thus, $I_{2}(z_{f})$ tends to one in the case when $\Omega _{0}>>\Delta _{0}$
and the light is completely transferred between the waveguides. We note that
Eq. (\ref{complete light transfer}) is valid not only if the coupling is
given as hyperbolic-secant shape, but apply universally to every symmetric
pulse-shaped smooth coupling ($\Omega (z)=\Omega (-z)$) that fulfills $%
\Omega (z_{i})=\Omega (z_{f})=0$ together with a sign flip of the phase
mismatch at the coupling maximum. An example of complete adiabatic light
switching between the two waveguides is shown in Figs. \ref{Fig2} and \ref%
{Fig3}. In the simulations shown in Fig. \ref{Fig2} and Fig. \ref{Fig3} we
have assumed hyperbolic-secant couplings, but any other smooth pulse-shaped
coupling may be used. The contour plot in Fig. \ref{Fig3} demonstrates the
robustness of the CLT against parameter variations.

\section{Achromatic beam splitter}


In this section we consider a symmetric array consisting of three coupled
optical waveguides, as shown in Fig. 4. We assume that the middle waveguide
is equally coupled to the two outer ones with coupling strength $\Omega (z)$
which is a function of the longitudinal coordinate $z$. Furthermore, the
propagation coefficients of the outer waveguides are assumed to be equal,
that is, both have an equal refractive index $n_{3}$ and hence equal
propagation coefficients $\beta _{1}$, while the middle waveguide's
refractive index changes from $n_{2}$ to $n_{4}$ at the maximum of the
coupling which also changes it's propagation coefficient $\beta _{2}$. The
evolution of the light propagating in such a waveguide array is described
by,
\begin{equation}
i\dfrac{d}{dz}\left[
\begin{array}{c}
c_{1}(z) \\
c_{2}(z) \\
c_{3}(z)%
\end{array}%
\right] =\left[
\begin{array}{ccc}
0 & \Omega (z) & 0 \\
\Omega (z) & \Delta (z) & \Omega (z) \\
0 & \Omega (z) & 0%
\end{array}%
\right] \left[
\begin{array}{c}
c_{1}(z) \\
c_{2}(z) \\
c_{3}(z)%
\end{array}%
\right] .  \label{three states}
\end{equation}%
where the phase mismatch $\Delta (z)=\beta _{1}-\beta _{2}$, $c_{(1,3)}(z)$
is the light amplitude in one of outer waveguides (the system is completely
symmetric), and $c_{2}(z)$ is the amplitude in the middle waveguide.

Notably, Eq. \eqref{three states} describing the light evolution in a system
of three evanescently coupled waveguides is analogous to the Schr\"{o}dinger
equation describing a three-state quantum system subject to external
electromagnetic field. It is well-known that the Hamiltonian of Eq. %
\eqref{three states} has a zero eigenvalue whose eigenvector is a so-called
\textquotedblleft dark state\textquotedblright\ of the system, that is, it
does not evolve under the evolution described by the Hamiltonian \cite%
{Vitanov1998}. We introduce a new basis states including the dark state
amplitude $c_{d}$ using the transformation.
\begin{equation}
\left[
\begin{array}{c}
c_{1}(z) \\
c_{2}(z) \\
c_{3}(z)%
\end{array}%
\right] =\left[
\begin{array}{ccc}
1/\sqrt{2} & 0 & 1/\sqrt{2} \\
0 & 1 & 0 \\
1/\sqrt{2} & 0 & -1/\sqrt{2}%
\end{array}%
\right] \left[
\begin{array}{c}
c_{b}(z) \\
c_{2}(z) \\
c_{d}(z)%
\end{array}%
\right] ,
\end{equation}%
where $c_{2}$ is the unchanged amplitude of the middle waveguide, and $c_{b}$
stands for the \textquotedblleft bright\textquotedblright\ equal
superposition of the amplitudes $c_{(1,3)}$. Rewriting Eq. (\ref{three
states}) in the new basis we obtain
\begin{equation}
i\dfrac{d}{dz}\left[
\begin{array}{c}
c_{b}(z) \\
c_{2}(z) \\
c_{d}(z)%
\end{array}%
\right] =\left[
\begin{array}{ccc}
0 & \Omega (z)\sqrt{2} & 0 \\
\Omega (z)\sqrt{2} & \Delta (z) & 0 \\
0 & 0 & 0%
\end{array}%
\right] \left[
\begin{array}{c}
c_{b}(z) \\
c_{2}(z) \\
c_{d}(z)%
\end{array}%
\right] .  \label{beam splitter}
\end{equation}%
Indeed, we find that the state $c_{d}$ is decoupled from states $c_{b}$ and $%
c_{2}$ and the three-state problem is reduced to a two-state one involving
states $c_{b}$ and $c_{2}$ only.

\begin{figure}[tbh]
\centerline{\includegraphics[width=0.7\columnwidth]{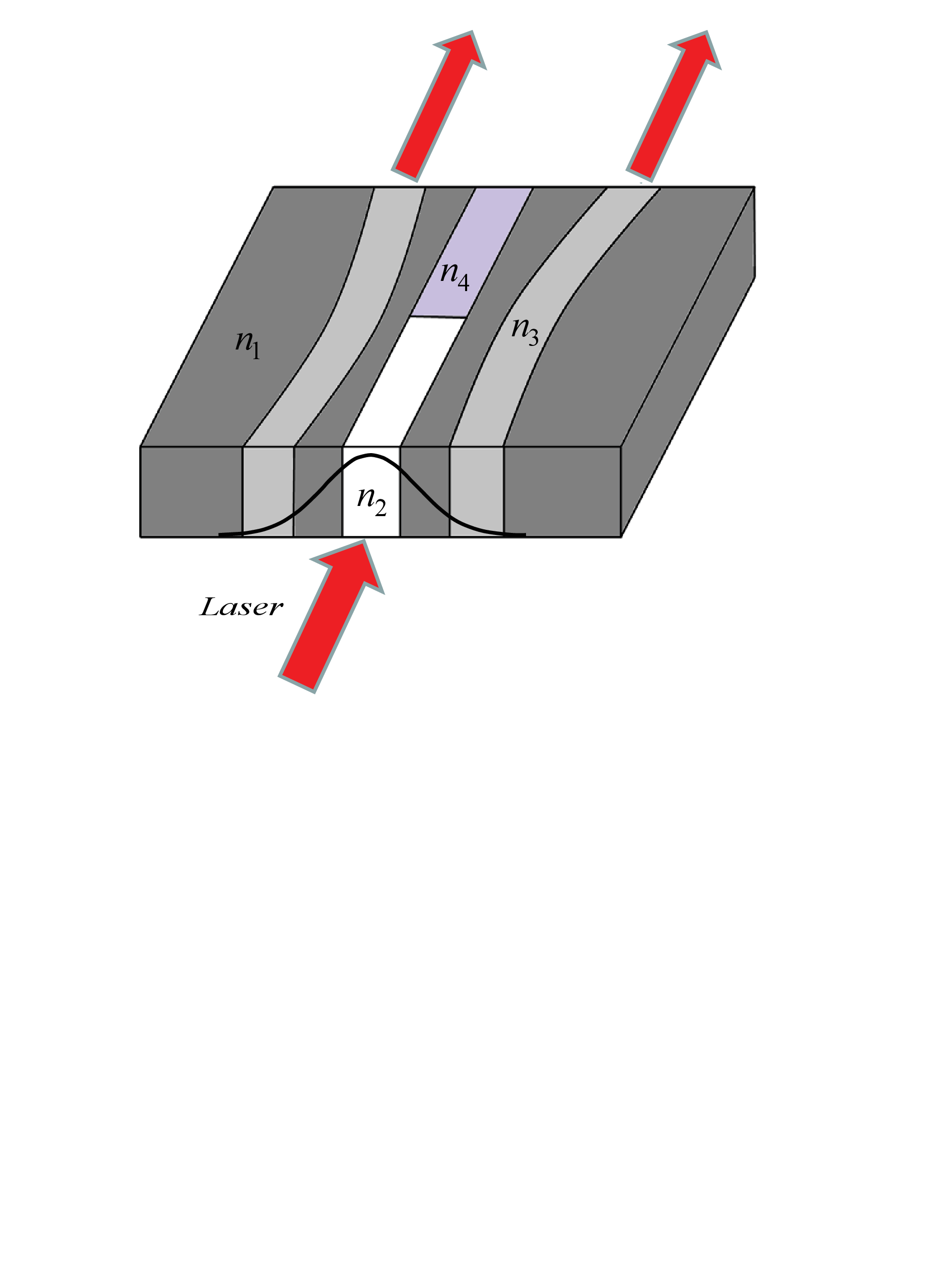}}
\caption{(Color online) Evanescently coupled three waveguides made of three
slabs with refractive indexes $n_{2}$, $n_{3}$ and $n_{4}$, embedded in a
medium of refraction index $n_{1}$. Gaussian shaped beam light is injected
initially in the middle waveguide and at the end it is achromatically
divided equally between the right and the left waveguide.}
\label{Fig4}
\end{figure}

In order to realize an achromatic beam splitter we take the following steps.
Initially, we input the light in the middle waveguide with state amplitude $%
c_{2}$ and following the evolution described by Eq. \eqref{beam splitter}
which is similar to that described in Sections \ref{Sec3} and \ref{Sec4},
the light is completely and robustly transferred into state $c_{b}$, which
is an equal superposition of the states of waveguides 1 and 3. Thus, the
light at the end of the waveguides will be split equally between waveguides
1 and 3 (outer waveguides) as shown in Fig. \ref{Fig4}. The light switching,
as shown on Fig. \ref{Fig3}, is robust against variations in the coupling $%
\Omega (z)$ and phase mismatch $\Delta (z)$; therefore, the technique is
expected to be achromatic. In contrast to previously suggested achromatic
adiabatic multiple beam splitters, which are based on an analog of
stimulated Raman adiabatic passage from quantum optics and are
unidirectional \cite{Dreisow,Ciret12,Rangelov}, the above proposed beam
splitting device works in forward and backward directions of light
propagation equally well. Hence, the above described achromatic beam
splitter is also bidirectional.


\section{Conclusions}

\label{conclusion}

In conclusion, we presented a two waveguide coupler configuration which
realizes complete achromatic all-optical switching robust to parameter
fluctuations. We showed that the light propagation in the proposed waveguide
coupler has an exact analytic solution which has the advantage of being
valid for any values of the interaction parameters. In the limit of large
coupling, complete light switching is achieved, which is insensitive to
parameter fluctuations and is therefore achromatic. We furthermore showed
that such a waveguide coupler can also be used for complete adiabatic light
switching. An extension of this system to three coupled planar waveguides
can be used as an achromatic beam splitter. We note that the achromaticity
of the light transfer is guaranteed by the adiabatic nature of the process.
Finally, the proposed waveguides coupler and beam splitter are
experimentally feasible using photoinduced reconfigurable planar waveguides.
The shapes and constants of propagation of such waveguides can be freely
controlled by changing the local refractive index of the crystal with
illuminating control light \cite{Dittrich,Gorram,Ciret12,Ciret13}.


\section*{Acknowledgements}

We acknowledge financial support by SUTD start-up Grant No. SRG-EPD-2012-029
and SUTD-MIT International Design Centre (IDC) Grant No. IDG31300102.

\end{document}